\documentclass[12pt`]{article}
\newcommand  {\Rbar} {{\mbox{\rm$\mbox{I}\!\mbox{R}$}}}

\newcommand  {\Nbar} {{\mbox{\rm$\mbox{I}\!\mbox{N}$}}}
\newcommand{\Cox}{{\hspace*{\fill}\rule{2mm}{2mm}\linebreak}}

\newsavebox{\zzzbar}
\sbox{\zzzbar}
  {\setlength{\unitlength}{0.9em}
  \begin{picture}(0.6,0.7)
  \thinlines
  \put(0,0){\line(1,0){0.6}}
  \put(0,0.75){\line(1,0){0.575}}
  \multiput(0,0)(0.0125,0.025){30}{\rule{0.3pt}{0.3pt}}
  \multiput(0.2,0)(0.0125,0.025){30}{\rule{0.3pt}{0.3pt}}
  \put(0,0.75){\line(0,-1){0.15}}
  \put(0.015,0.75){\line(0,-1){0.1}}
  \put(0.03,0.75){\line(0,-1){0.075}}
  \put(0.045,0.75){\line(0,-1){0.05}}
  \put(0.05,0.75){\line(0,-1){0.025}}
  \put(0.6,0){\line(0,1){0.15}}
  \put(0.585,0){\line(0,1){0.1}}
  \put(0.57,0){\line(0,1){0.075}}
  \put(0.555,0){\line(0,1){0.05}}
  \put(0.55,0){\line(0,1){0.025}}
  \end{picture}}
\newcommand{\Zbar}{\mathord{\!{\usebox{\zzzbar}}}}

\newtheorem{prop}{Proposition}[section]
\newtheorem{thm}{Theorem}[section]

\newtheorem{example}{Example}[section]
\newcommand{\Z}{\Zbar}
\newcommand{\R}{\Rbar}
\newcommand{\N}{\Nbar}

\newcommand{\s}{\ensuremath{\mathcal{S} }}

\newcommand{\F}{\ensuremath{\mathcal{F} }}

\begin{document}

\setlength{\textheight}{21cm}

\title{{\bf The Fluctuation Theorem as a Gibbs Property}}

\author{{\bf Christian
Maes}\thanks{Onderzoeksleider FWO, Flanders. \ Email:
christian.maes@fys.kuleuven.ac.be } , Instituut voor  Theoretische
Fysica, \\ K.U.Leuven, B-3001 Leuven, Belgium.}

\maketitle

Dedicated to the memory of Edwin T. Jaynes.
\begin{abstract}
Common ground to recent studies exploiting relations between
dynamical systems and non-equilibrium statistical mechanics is, so
we argue, the standard Gibbs formalism applied on the level of
space-time histories. The assumptions
(chaoticity principle) underlying the Gallavotti-Cohen fluctuation
theorem make it possible, using symbolic dynamics, to employ the
theory of one-dimensional lattice spin systems.  The Kurchan and
Lebowitz-Spohn analysis of this fluctuation theorem for stochastic
dynamics can be restated on the level of the space-time measure
which is a Gibbs measure for an interaction determined by the
transition probabilities. In this note we understand the
fluctuation theorem as a Gibbs property as it follows from the very
definition of Gibbs state.  We give a local version of the
fluctuation theorem in the Gibbsian context and we derive from this
a version also for some class of spatially extended stochastic
dynamics.
\end{abstract}
\vspace{3mm}

\section{Context and main observations.}

\subsection{Scope.}
The fluctuation theorem of Gallavotti and Cohen, see
\cite{gc,gc1,Ru4}, asserts that for a class of dynamical systems
the fluctuations in time of the phase space contraction rate obey a
general law.  We refer to the cited literature for additional
details and precision and we only sketch here the main ingredients.
\\ One considers a reversible smooth dynamical system $\xi\rightarrow
\phi(\xi),\xi \in \Omega$. The phase space $\Omega$ is
in some sense bounded carrying only a finite number of degrees of
freedom (a compact and connected manifold). The transformation
$\phi$ is a diffeomorphism of $\Omega$.  The resulting (discrete)
time evolution is obtained by iteration and the reversibility means
that there is a diffeomorphism $\theta$ on $\Omega$ with
$\theta^2=1$ and $\theta\circ
\phi
\circ \theta = \phi^{-1}$. Consider now minus the logarithm of the Jacobian
 determinant $J$ which arises from the change of variables implied by the dynamics.
We write $\dot{S} \equiv - \log J$.  One is interested in the
fluctuations of
\begin{equation}\label{gccur}
w_N(\xi) \equiv \frac 1 {\rho(\dot{S}) N} \sum_{-N/2}^{N/2}
\dot{S}(\phi^n(\xi)),
\end{equation}
for large time $N$. Here, $\rho$ is the stationary probability
measure (SRB measure) of the dynamics with expectations
\begin{equation}\label{srb}
\rho(f) = \lim_N \frac 1{N} \sum_0^N f(\phi^n \xi)
\end{equation}
corresponding to time-averages for almost every randomly chosen
initial point $\xi\in \Omega$.  This random choice refers to an
absolutely continuous measure with respect to the Riemann volume
element $d\xi$ on $\Omega$ (and is thought of as describing the
microcanonical ensemble for $\Omega$ the energy surface).
$\dot{S}(\xi)$ is the phase space contraction rate (which is
identified with the entropy production rate) and one assumes (and
sometimes proves) dissipativity:
\begin{equation}\label{popo}
\rho(\dot{S}) > 0.
\end{equation}
It is assumed that the dynamical system satisfies some technical
(ergodic) condition: it is a transitive Anosov system. This ensures
that the system allows a Markov partition (and the representation
via some symbolic dynamics) and the existence of the SRB measure
$\rho$ in (\ref{srb}). This technical assumption is not taken
physically very serious but instead it is supposed to guide us
towards general results which are true in a broader
context\footnote{The situation resembles here to some extent that
for the ergodic hypothesis.  Ergodicity is likely to be false in
quite a number of realistic situations and in any event it is
irrelevant. Nevertheless assuming ergodicity can lead to correct
consequences.}. That is what is affirmed in the so called chaotic
hypothesis: ``A reversible many particle system in a stationary
state can be regarded as a transitive Anosov system for the purpose
of computing the macroscopic properties,'' see also e.g.
\cite{gc,gc1,G,G1,Ru3,G2}. The fluctuation theorem then states that
$w_N(\xi)$ has a distribution $\rho_N(w)$ with respect to the
stationary state $\rho$ such that
\begin{equation}\label{flucthm}
\lim_N \frac 1{N \rho(\dot{S})w}\ln \frac{\rho_N(w)}{\rho_N(-w)} = 1
\end{equation}
always.  In other words, the distribution of entropy production
over long time intervals satisfies some general symmetry
property.\\ This theorem originated from numerical evidence, e.g.
in \cite{ECM}, and it has various interesting consequences. For
example, in \cite{G}, it was interpreted as extending the
Green-Kubo formulas to arbitrary forcing fields for a class  of
non-equilibrium dynamics.

In \cite{K}, Kurchan pointed out that this fluctuation theorem
also holds for certain diffusion processes.  This is the context of
finite systems undergoing Langevin dynamics.  This was extended by
Lebowitz and Spohn in \cite{LS} to quite general Markov processes.
There was however no general scheme for identifying the quantity
(being some analogue of (\ref{gccur})) for which the fluctuation
theorem holds. Yet, from applying the fluctuation theorem in this
context to simple models of stochastic dynamics, relations appeared
between the entropy production and the action functional satisfying
the theorem.

In this note we understand the fluctuation theorem within the Gibbs
formalism.  Since this formalism is often considered as giving a
mathematical structure to the theory of equilibrium statistical
mechanics and in order to avoid misunderstanding, we insist from
the beginning that we wish to see this Gibbs formalism applied here
to nonequilibrium conditions.  The right way of looking at it, is
to consider space-time histories drawn from a Gibbs measure.  In
other words, our analysis is not to be regarded as an investigation
of fluctuations in an equilibrium system or as the restriction of
the fluctuation theorem to equilibrium conditions. On the contrary,
the observations we make can be seen as underlying and (at least in
some sense) extending both the Gallavotti-Cohen and the Kurchan and
Lebowitz-Spohn fluctuation theorems. Underlying because the
technical conditions of the Gallavotti-Cohen work reduce to a large
extent the fluctuation theorem to a statement about one-dimensional
Gibbs measures. That is not very different in the Lebowitz-Spohn
work where the strong chaoticity is replaced by stochasticity and
the Perron-Frobenius theorem is applied to the dynamical generator
as it is usually done for the transfer matrix in one-dimensional
Gibbs states. The fact that something more general and typical of
Gibbs states is at work here was already announced in Section 3 of
\cite{BGG} where Example \ref{ex2} below was applied to the
one-dimensional Ising model in an external field.  Our work
systematizes this remark. But for Gibbs states, the fluctuation
theorem does not rely on having one dimension or on a high
temperature condition.  Once this is perceived, one is tempted to
conclude that chaoticity assumptions, while important guides,
cannot really be necessary for a fluctuation theorem or its
consequences to hold.  Perhaps it is more natural to assume
immediately that for the purpose of computing macroscopic
properties, a many particle system in a steady state should be
regarded as a Gibbs system for the space-time histories.  And we
know that the reason for Gibbs distributions has little to do with
the detailed properties of the system's dynamics but instead is
based on  statistical principles \footnote{We have in mind the
maximum entropy principle and the foundations of statistical
mechanics in the theory of large deviations, see e.g.
\cite{Lan,Ja}. This must be contrasted with the approach from the
theory of dynamical systems (as summarized for example in
\cite{Ru4}).  Notice that Markov partitions do not correspond to a
statistical procedure; they fully encode the dynamics.}. There is
finally a second, practically speaking, more important extension of
the earlier results.  In the analysis below, we present a local
version of the fluctuation theorem. A mechanism for the validity of
a local version was already discussed in the recent \cite{G3}. This
is crucial because it is only a local fluctuation theorem that
leads to observable consequences and we will see that this is quite
natural in a Gibbsian setup.

\subsection{Disclaimers.}

Our analysis below is limited in various ways including:\\
\noindent
1. Time (and space) is discrete: a regular lattice plays the role
of space-time.  We believe that going to continuous time is a
technical step (which is not expected to be very difficult) and
that this is irrelevant for the purpose of the paper.\\
\noindent
2. No hard-core conditions: we take a smooth potential and all
transition probabilities are bounded away from zero.  In
particular, this seems problematic when dealing with dynamics
subject to certain conservation laws.  Again, we do not think that
this is essential because the Gibbs properties we use also hold for
hard-core interactions. Extra care and conditions would be needed
for writing down certain formulae but we believe they do not modify
the main result.\\
\noindent
3. Discrete spins: we deal with regular lattice spin systems. While
some compactness of phase space is nice to have around, our results
depend solely on having a large deviation principle for Gibbs
states.  The extent to which such a principle holds decides on the
possible extensions of our results.\\
\noindent
4. No phase transitions: while the fluctuation theorem holds quite
generally, its contents can be empty when the large deviations
happen on another scale than linear in time (and spatial volume).
 In other words, the corresponding rate function could fail to be strictly convex
 in which case instabilities or phase transitions are present.
These `violations' of the fluctuation theorem can of course not
happen when the spatial volume is finite (for a sufficiently
chaotic dynamics or for a non-degenerated stochastic dynamics) or,
for infinite systems, when we are in the `high temperature' regime.
Such scenario's are of course well-documented for Gibbs states.\\
\noindent
5. Steady states and time-homogeneity: we do not consider here the
(physically very relevant) problem of forces or potentials
depending on time nor do we investigate here the long time behavior
of the system started in anything other than in a stationary state.
In these cases, we must refer to the study of Gibbs states on
half-spaces with particular boundary conditions but the main points
must remain intact.\\
We hope to include in a future publication the extensions mentioned
above.  In particular, all examples that appear in \cite{LS} can be
systematically obtained using the one and same algorithm that will
be explained below.  We will briefly illustrate such a result (for
a continuous time dynamics with a conservation law) at the end
(Section 3.3).

\subsection{Notation and definitions.}

We restrict ourselves here to lattice spin systems. For lattice we
take the regular $d+1$-dimensional set $\Z^{d+1}, d\geq 0$. The
reason for taking $d+1$ is that the extra dimension refers to the
time-axis.  The points of the lattice are denoted by $x,y,\ldots$
with $x=(i,n), n \in \Z, i \in
\Z^d$.  We can read the time by the mapping $t(x) = n$ if $x=(i,n)$.
The distance between two points $x=(i,n),y=(j,m)\in \Z^{d+1}$ is
$|x-y|
\equiv
\max\{|n-m|,|i-j|\}$ with $|i-j|\equiv
\max\{|i_1-j_1|,\ldots,|i_d-j_d|\}$ for the two sites $i =
(i_1,\ldots,i_d), j=(j_1,\ldots,j_d) \in \Z^d$.
The set of finite and non-empty subsets of $\Z^{d+1}$ is
denoted by $\cal S$.  For general elements of $\cal S$ we write
$\Lambda, A,\ldots$;
they correspond to (finite) space-time regions.  $\Lambda^c = \Z^{d+1} \setminus
\Lambda$ is the complement
of $\Lambda$; $|\Lambda|$ is the cardinality of $\Lambda$.

A space-time configuration of our lattice spin system is denoted by
$\sigma,\eta,\xi,\ldots$.  This is a mapping $\sigma: \Z^{d+1}
\rightarrow S$ with values $\sigma(x) \in S$ in the single site
state space $S$ which is taken finite.  Ising spins have
$S=\{+1,-1\}$.  The set of all configurations is $\Omega_{d+1} =
S^{\Z^{d+1}}$. By $\sigma_E, E\subset \Z^{d+1}$ we mean, depending
on the situation, both the restriction of $\sigma$ to $E$ as well
as a configuration on $E$, i.e. an element of $S^E$.  The
configuration $\sigma_\Lambda\eta_{\Lambda^c}$ is equal to $\sigma$
on $\Lambda$ and is equal to $\eta$ on $\Lambda^c$.\\
$\Omega_{d+1}$ is equipped with the product topology and is a
compact space.  If we denote by ${\cal F}_o$ the set of all subsets
of $S$, then ${\cal F}_\Lambda = {\cal F}_o^\Lambda$ is the Borel
sigma-algebra generated by the $(\sigma(x), x\in \Lambda)$.  We
write ${\cal F} \equiv {\cal F}_{\Z^{d+1}}$; $(\Omega,{\cal F})$ is
the measurable space of space-time
 configurations.\\
 Local functions on $\Omega_{d+1}$ are
real-valued functions $f$ which are ${\cal F}_\Lambda$-measurable
for some $\Lambda \in {\cal S}$.  The (finite) dependence set of such
a local $f$ is denoted by $D_f$. A continuous function is
every function on $\Omega_{d+1}$ which is the uniform limit of local
functions.  The uniform norm is denoted by $\|f\| \equiv
\sup_\sigma |f(\sigma)|$.\\
Finally, configurations and functions on $\Omega_{d+1}$ can be
translated
over $\tau_x, x\in \Z^{d+1}: \tau_x \sigma(y) = \sigma(y+x),
  \tau_x f(\sigma)=f(\tau_x \sigma)$. If clear from the context, we
also write $f_x$ for $\tau_x f$.\\

 We consider families of local one-to-one (invertible) transformations
$\pi_\Lambda$ on $S^\Lambda$ where $\Lambda$ will vary in some
large enough subset of $\cal S$ (which will be specified later on). As
maps on $\Omega_{d+1}$ they have the properties that
\begin{enumerate}\label{trans}
\item
\begin{equation}\label{pi1}
\pi_\Lambda(\sigma)(x) = \sigma(x), x\in \Lambda^c;
\end{equation}
\item
\begin{equation}\label{pi2}
\pi_\Lambda \circ \pi_\Lambda = 1.
\end{equation}
\item
\begin{equation}
\pi_\Lambda \circ \tau_x = \pi_{\Lambda+x}
\end{equation}
\item
\begin{equation}
\pi_\Lambda(\sigma)(x) = \pi_{\Lambda'}(\sigma)(x)
\end{equation}
for all $x\in \Lambda \subset \Lambda'$.
\end{enumerate}
For every function $f$ on $\Omega_{d+1}$, we write $\pi_\Lambda
f(\sigma)
\equiv f(\pi_\Lambda(\sigma))$. The (product over $\Lambda$ of the)
counting measure
on $S^\Lambda$ is invariant under $\pi_\Lambda$. Notice that the function
$\Delta_{\pi_\Lambda}f \equiv \pi_\Lambda f - f$ satisfies
$\pi_\Lambda \Delta_{\pi_\Lambda}f = -\Delta_{\pi_\Lambda}f$. We
give two interesting examples of such a transformation.
\begin{example}\label{ex1}
Take $\Lambda=\Lambda_{N,L}$ a rectangular shaped region
centered at
the origin with time-extension $2N+1$ and spatial volume $(2L+1)^d$.
The transformation
$\pi_\Lambda(\sigma)(j,m)=\sigma(j,-m), |j|\leq L, |m|\leq N$
time-reverses the space-time configuration in the window
$\Lambda_{N,L}$.
\end{example}
\begin{example}\label{ex2}
Take the Ising-case $S=\{+1,-1\}$ and $\pi_\Lambda(\sigma)(x)=
-\sigma(x), x\in \Lambda$ corresponding to a spin-flip in
$\Lambda\in {\cal S}$.
\end{example}

Probability measures on $(\Omega_{d+1},\cal F)$ are denoted by
$\mu,\nu,\ldots$.  The corresponding random field is written as
$X=(X(x), x\in \Z^{d+1})$. The expectation of a function $f$ is
written
as $\mu(f) \equiv \int f(\sigma) \mu(d\sigma)$.  As {\it a priori}
measure we take the uniform product measure $d\sigma$ with normalized
counting measure as marginals, for which $\int f(\sigma) d\sigma
\equiv 1/|S|^{|\Lambda|}
\sum_{\sigma_\Lambda} f(\sigma_\Lambda)$ when $f$ is ${\cal
F}_\Lambda$-measurable.\\ We will be dealing with Gibbs states
$\mu$ in what follows; $\mu$ is a Gibbs measure with respect to the
Hamiltonian $H$ at inverse temperature $\beta$ (and always with
respect to the counting measure as {\it a priori} measure) when for
every $\Lambda \in {\cal S}$ and for each pair of configurations
$\sigma_\Lambda, \eta_\Lambda \in S^{\Lambda}$
\begin{equation}\label{gibbs}
\frac{\mu[X(x) = \sigma(x),x\in \Lambda|X(x)=\xi(x), x\in
\Lambda^c]}
{\mu[X(x) = \eta(x),x\in \Lambda|X(x)=\xi(x), x\in
\Lambda^c]} =
\exp[-\beta(H(\sigma_\Lambda\xi_{\Lambda^c})-H(\eta_\Lambda\xi_{%
\Lambda^c}) ) ]
\end{equation}
for $\mu-$ almost every $\xi \in \Omega_{d+1}$.   The Hamiltonian
$H = \sum_A U_A$ is formally written as a sum of (interaction)
potentials $U_A(\sigma) = U_A(\sigma_A)$ with well-defined relative
energies $H(\sigma) - H(\eta)$ for $\{x\in \Z^{d+1}:
\sigma(x)\neq\eta(x)\} \in {\cal S}$ if $\sum_{A\ni x} \|U_A\| <
\infty, x\in Z^{d+1}$.  Other weaker conditions than uniform
absolute summability of the potential are possible.  The essential
Gibbs property is (\ref{gibbs}) which identifies the existence of a
well-defined relative energy governing the relative weights of
configurations that locally differ.  (\ref{gibbs}) is the infinite
volume version of the equivalent statement for finite volume Gibbs
states
\begin{equation}\label{pfie}
\mu_\Lambda(\sigma_\Lambda|\eta_{\Lambda^c}) = \frac
1{Z_\Lambda^\beta(\eta)} \exp[-\beta\sum_{A \cap \Lambda \neq
\emptyset} U_A(\sigma_\Lambda \eta_{\Lambda^c})],
\end{equation}
with $Z_\Lambda^\beta(\eta)$ the normalizing factor (partition
function with $\eta$ boundary conditions).\\ Traditionally, Gibbs
measures give the distribution of the microscopic degrees of
freedom for a macroscopic system in thermodynamic equilibrium. The
choice of the ensemble is determined by the experimental situation
and is fixed by the choice of the relevant macro-variables. There
is however no {\it a priori} reason to exclude nonequilibrium
situations from the Gibbs formalism if considered as a procedure of
statistical inference.  Then the information concerning the
nonequilibrium state (like obtained from measuring the currents) is
incorporated in the ensemble.  Moreover, as we will use in Section
3, one can in many cases explicitly construct the Gibbs states
governing the space-time distribution as the path-space measure for
the dynamics.  The fact that these examples concern stochastic
dynamics should not be regarded as a return to the strongly chaotic
regime but rather as the proper way to deal with incomplete
knowledge about the microscopic configuration of a system composed
of a huge number of locally interacting components.

\subsection{Main observation.}

We start with the simplest observation.  The rest will
follow as immediate generalizations
(with perhaps a slightly more complicated notation).\\
Look at (\ref{gibbs}).  This Gibbs property implies that the image
measure of $\mu$ under a transformation that affects only the spins in
$\Lambda$ is absolutely continuous with respect to $\mu$ with the
Boltzmann-Gibbs factor as Radon-Nikodym derivative.  Putting it
simpler, it is an immediate consequence of the Gibbs property that
for all continuous functions $f$
\begin{equation}\label{wicht}
\mu(\pi_\Lambda f) = \mu( f W_\Lambda)
\end{equation}
with $W_\Lambda \equiv \exp[-\beta \sum_{A\cap \Lambda\neq
\emptyset}(\pi_\Lambda U_A - U_A)]$.  But now the road is straight:
take $f = W_\Lambda^{\lambda-1}$ in (\ref{wicht}) and compute
\begin{equation}
\mu(W_\Lambda^\lambda) = \mu(W_\Lambda^{\lambda-1} W_\Lambda).
\end{equation}
From (\ref{wicht}) this is equal to
\begin{equation}
\mu(\pi_\Lambda W_\Lambda^{\lambda-1}) = \mu(W_\Lambda^{1-\lambda})
\end{equation}
where the last equality follows from $\pi_\Lambda
(W_\Lambda^{\lambda-1}) = W_\Lambda^{1-\lambda}$. Thus, it is
immediate that Gibbs measures satisfy
\begin{equation}\label{1.14}
\mu(e^{-\lambda \beta R_\Lambda}) =
\mu(e^{-(1-\lambda)\beta R_\Lambda}), \lambda\in \R
\end{equation}
with relative energy
$R_\Lambda\equiv\pi_\Lambda H - H$ corresponding to
the transformation $\pi_\Lambda$:
\begin{equation}\label{relha}
R_\Lambda = \sum_{A\cap\Lambda\neq \emptyset}[\pi_\Lambda U_A -
U_A].
\end{equation}

We now imagine the above for a sequence of volumes
$\Lambda$ growing
 to $\Z^{d+1}$ in a sufficiently regular manner (e.g. increasing cubes).
 Suppose now furthermore that $\mu$ is a Gibbs measure
for a translation-invariant interaction potential and that
\begin{equation}
R_\Lambda(\sigma)  =
\sum_{x\in \Lambda} \tau_x J(\sigma) + h_\Lambda(\sigma)
\end{equation}
with $J$ a bounded continuous function and $||h_\Lambda||/|\Lambda|
\rightarrow 0$ as $\Lambda$ becomes infinite. This will be made explicit later on.
 Then, the following
limit exists:
\begin{equation}
p(\lambda J|\mu) \equiv
- \lim_{\Lambda} \frac 1{|\Lambda|} \ln \mu[ \exp(-\beta \lambda
\sum_{x\in \Lambda} J_x)] \end{equation}
with $J_x \equiv \tau_x J$, and, from (\ref{1.14}), it satisfies
\begin{equation}
p(\lambda J|\mu) = p((1-\lambda)J|\mu).
\end{equation}
As a consequence, its Legendre transform
\begin{equation}
i_J(w|\mu) \equiv \sup_\lambda[p(\lambda J|\mu) - \lambda w]
\end{equation}
satisfies
\begin{equation}\label{cg}
i_J(w|\mu) - i_J(-w|\mu) = -w.
\end{equation}
It is not necessary (but it is possible) to employ the whole machinery of
the theory of large
deviations for Gibbs states to understand what this means: the probability
law $P_\Lambda(w)$ for the random variable $\sum_{x\in \Lambda}
J_x(X)/|\Lambda|$
as induced from the random field $(X(x), x\in \Z^{d+1})$ with distribution
$\mu$, behaves (for large $\Lambda$) as
\begin{equation}\label{ldev}
P_\Lambda(w) \sim e^{-i_J(w|\mu) |\Lambda|}.
\end{equation}
and the rate function $i_J(w|\mu)$ satisfies (\ref{cg}). Comparing
this with (\ref{flucthm}), we see we have obtained exactly the same
structure as in the Gallavotti-Cohen fluctuation theorem with
practically no effort.

\subsection{Plan.}

We first present the fluctuation theorem in a Gibbsian context
without too much reference to an underlying dynamics through which,
possibly, the Gibbs states are obtained as space-time measures.
Yet, to avoid misunderstanding, we repeat that we think of these
Gibbs measures here as describing the steady states or symbolic
dynamics for some spatially-extended non-equilibrium dynamics. They
are to be thought of as distributions for the space-time histories.
Via standard thermodynamic relations, we give the relation between
the action functional satisfying the large deviation principle
(fluctuation theorem) and the relative entropy between the forward
and the backward evolution. In particular, in quadratic
approximation, the Green-Kubo formula appears. Time enters
explicitly in Section 3 where via the example of probabilistic
cellular automata the general philosophy is illustrated.

\section{Fluctuation theorem for Gibbs states.}

In the present setup, we have no {\it a priori} reason to prefer
one lattice direction over another and we fix the family of
increasing cubes $\Lambda_n$ of side length $n\in \N_0$ centered
around the origin in which we are going to apply the
transformations $\pi_{\Lambda_n}\equiv
\pi_n$ having the properties described in Section \ref{trans}.  For
every $A\in {\s}$ we write $A_n$ for the smallest cube $\Lambda_n$
(with $n=n(A)$) for which $A\subset
\Lambda_n$.

\subsection{Symmetry breaking potential.}

In what follows we simply set $\Omega=\Omega_{d+1}$. A potential
$U$ is a real-valued function on ${\s} \times
\Omega$ such that $U_A \in {\F}_A$ (i.e. only depending on the spins inside $A$)
for each $A\in {\s}$ (put $U_\emptyset \equiv 0$). It describes the
interaction between the spins in the region $A$. We consider a
family of $m+1$ interaction potentials $(U_A^\alpha)_A,
\alpha=0,\ldots,m$.  We assume translation-invariance, meaning
that
\begin{equation}\label{tra}
U_A^\alpha(\eta) = U_{A+x}^\alpha(\tau_x \eta),
\end{equation}
for all $A\in {\s},x \in \Z^{d+1},\eta \in \Omega$. As usual we
also take it that the total interaction of a finite region with the
rest of the lattice is finite, i.e. we assume that the potential is
uniformly absolutely summable:
\begin{equation}\label{abs}
\sum_{A\ni 0} \|U_A^\alpha\| < \infty.
\end{equation}
(This assumption of uniformity is not strictly needed but it avoids
irrelevant technicalities.  Similarly, hard core interactions are
also not excluded but extra care and assumptions would be needed.)
Given the family of transformations $\pi_n$, we define the relative
energies
\begin{equation}\label{rele}
R_n^\alpha \equiv \sum_{A\cap\Lambda_n\neq \emptyset} (\pi_n
U_A^\alpha
- U_A^\alpha), \alpha=0,\ldots,m.
\end{equation}
We make a difference between the potential $U^0$ and the $U^\alpha,
\alpha=1,\ldots,m$ from their behavior under the $\pi_n$.
We assume that $U^0$ is invariant under the $\pi_n$ in the sense
that $\pi_n U_A^0 = U_A^0$ whenever $n\geq n(A)$ implying that
\begin{equation}
\lim_n \frac{\|R^0_n\|}{|\Lambda_n|} = 0.
\end{equation}
The reason for taking $m>1$ is to allow for and to distinguish
between possibly different mechanisms for breaking the symmetry of
the reference interaction $U^0$.

We define the current associated to the symmetry breaking
interaction $U^\alpha,\alpha=1,\ldots,m$ to be
\begin{equation}\label{cur}
J_x^\alpha \equiv \lim_n \sum_{x\in A\subset\Lambda_n} \frac
1{2|A|}(\pi_{n(A)}U_A^\alpha - U_A^\alpha).
\end{equation}
$J_x^\alpha$ is a continuous function on $\Omega$ and, from
(\ref{tra}), $J_x^\alpha(\eta)=J_0^\alpha(\tau_x \eta)$. The term
`current' is suggestive for interpreting (\ref{cur}) as the real
current at the space-time point $x$ associated to some driving of a
reference steady state $\nu$ thereby breaking the
time-reversal symmetry in the case of Example \ref{ex1},
see next section. We take $\nu$ to be a Gibbs state with
respect to the interaction $U^0$, i.e. with formal Hamiltonian
\begin{equation}\label{h0}
H^0 \equiv \sum_A U_A^0,
\end{equation}
(see (\ref{gibbs})) for which the $\pi-$symmetry is unbroken:
\begin{equation}
\nu \circ \pi_n = \nu.
\end{equation}
As a consequence, the currents (\ref{cur}) vanish identically in
that state:
\begin{equation}
\nu(J_x^\alpha) = 0,\alpha=1,\ldots,m.
\end{equation}
The perturbed or driven state is denoted by $\mu$.  It is a
translation-invariant Gibbs state at inverse temperature $\beta$
with respect to the formal Hamiltonian
\begin{equation}
H\equiv H^0 + \sum_{\alpha=1}^m E^\alpha H^\alpha,
\end{equation}
where the $H^\alpha$ are built (as in (\ref{h0})) from the
interaction potentials $U^\alpha$ and where the $E^\alpha$ are real
numbers parameterizing the strength of a symmetry breaking or
driving force.  As before, in the definition of Gibbs states, we
always take the normalized counting measure as {\it a priori}
measure, see (\ref{gibbs}).

\subsection{Fluctuation theorem.}

\begin{thm}\label{thm:cg}
Suppose that $\mu$ is a translation-invariant Gibbs state
 for the translation-invariant potential
  $(U_A=U_A^0+\sum_{\alpha=1}^m E^\alpha U_A^\alpha)_A$
  as in the preceding subsection.
The limit
\begin{equation}\label{free}
p(\lambda,E) \equiv - \lim_n \frac 1{|\Lambda_n|} \ln
\mu[e^{-\beta\sum_{x\in\Lambda_n}
\sum_{\alpha=1}^m \lambda_\alpha J^{\alpha}_x}]
\end{equation}
exists and satisfies
\begin{equation}\label{flu}
p(\lambda,E) = p(2E-\lambda,E)
\end{equation}
for every $\lambda\equiv (\lambda_1,\ldots,\lambda_m)$ and $
E\equiv (E^1,\ldots,E^m)
\in \R^m$.
\end{thm}
\medskip
\noindent
{\bf Proof:} The existence of the limit is a standard result of the
Gibbs formalism, see e.g. \cite{Geo,EFS,Sim}.  As announced via
(\ref{wicht}) the main observation leading to (\ref{flu}) is that
\begin{equation}
\mu(\pi_n f) = \mu(\exp[-\beta\sum_{A\cap\Lambda_n\neq \emptyset}(\pi_n U_A - U_A)] f),
\end{equation}
simply because $\mu$ is a Gibbs state for the potential $(U_A)$ at
inverse temperature $\beta$.  Therefore, taking numbers
$h_\alpha,\alpha=1,\ldots,m$ and
$f=\exp[\beta\sum_{\alpha=1}^m (1-h_\alpha) E^\alpha R_n^\alpha]$
in that formula,
\begin{eqnarray}
&& \mu(\exp[- \beta R_n^0 - \beta \sum_{\alpha=1}^m h_\alpha
E^\alpha R_n^\alpha])=
\nonumber \\ &&
\mu(\exp[- \beta R_n^0 - \beta \sum_{\alpha=1}^m
E^\alpha R_n^\alpha] \exp[
\beta \sum_{\alpha=1}^m (1-h_\alpha) E^\alpha R_n^\alpha])= \nonumber \\
&&
\mu(\exp[- \beta \sum_{\alpha=1}^m (1-h_\alpha) E^\alpha R_n^\alpha]).
\label{toep}
\end{eqnarray}
Now,
\begin{equation}\label{waar}
R_n^\alpha=2\sum_{x\in\Lambda_n} J_x^\alpha - I_1 + I_2
\end{equation}
where both
\begin{equation}
I_1 \equiv \sum_{x\in \Lambda_n}\sum_{A\ni x, A\cap
\Lambda_n^c\neq \emptyset}
\frac 1{|A|}(\pi_{n(A)}U_A^\alpha - U_A^\alpha)
\end{equation}
and
\begin{equation}
I_2 \equiv \sum_{A\cap\Lambda_n,A\cap
\Lambda_n^c\neq\emptyset}(\pi_n U_A^\alpha - U_A^\alpha)
\end{equation}
are small of order $o(|\Lambda_n|)$ because of (\ref{abs}):
$||I_i||/|\Lambda_n|
\rightarrow 0$ as $n$ goes to infinity, $i=1,2$.
Upon inserting (\ref{waar}) into (\ref{toep}) and taking $h_\alpha
=
\lambda_\alpha/2E^\alpha$ (for $E^\alpha\neq 0$), we get
\begin{equation}
\frac 1{|\Lambda_n|} |\ln \frac{\mu(\exp[-\beta \sum_{x\in\Lambda_n}
\sum_{\alpha=1}^m \lambda_\alpha J^{\alpha}_x])}{
\mu(\exp[-\beta \sum_{x\in\Lambda_n}
\sum_{\alpha=1}^m (2E^\alpha-\lambda_\alpha) J^{\alpha}_x])}|
\end{equation}
going to zero as $n\uparrow \infty$.  This is exactly what was
needed. $\Cox$
\medskip

\noindent
{\bf Remark 1:} Gibbs states satisfy a large deviation principle, see
e.g. \cite{Lan} and \cite{EFS} for additional references.  As a
result, (\ref{flu}) implies
(\ref{ldev})-(\ref{cg}).  We do not add a more precise formulation
here.

\noindent
{\bf Remark 2:} Related to this, as is clear from the proof, the
essential property is that the functionals $\{ \log
\frac{d(\mu\circ \pi_{\Lambda})}{d\mu} : \Lambda \in {\cal S} \}$
satisfy a large deviation principle under $\mu$.  We speak about
the (somewhat more restricted) Gibbs property because, in all cases
we have in mind, the large deviations arise from Gibbsianness of
the random field.

\noindent
{\bf Remark 3:} The theorem above provides a {\bf local} version of
the fluctuation theorem since the measure $\mu$ lives on a much
larger (in fact, infinite) volume than the size of the observation
window $\Lambda_n$.  The relations (\ref{1.14}) and (\ref{toep})
are identities exactly verified for the finite volumes $\Lambda_n$.
This is similar to the local fluctuation theorem of \cite{G3}.
Notice also that the limit $p(\lambda,E)$ exists and remains
unchanged if instead of taking the sequence of cubes $\Lambda_n$ we
take volumes $\Lambda$ growing to $\Z^{d+1}$ in the van Hove sense,
see e.g. \cite{Geo,EFS,Sim}. This will be exploited in the next
section (Theorem \ref{thm:cgpca}) to separate time from the spatial
volume.

\noindent
{\bf Remark 4:} The fluctuation theorem is formulated here (and
elsewhere) on a volume-scale, anticipating large deviations which
are exponentially small in the volume, see (\ref{ldev}). This is
certainly the typical behavior at high temperatures. However, the
same reasoning of the proof above remains equally valid for other
--- less disordered --- regimes where the large deviations may happen
on another scale. As an example, suppose that
\begin{equation}
a(\lambda,E) \equiv- \lim_n \frac 1{n^{d}} \ln
\mu[e^{-\frac{\beta}{n}\sum_{x\in\Lambda_n}
\sum_{\alpha=1}^m \lambda_\alpha J^{\alpha}_x}].
\end{equation}
Then, remembering that $\Lambda_n \sim n^{d+1}$, it also satisfies
\begin{equation}\label{flu1}
a(\lambda,E) = a(2E-\lambda,E).
\end{equation}
Such a scaling is applied in the study of large deviations in the
phase coexistence regime where the probability of a droplet of the
wrong phase is only  exponentially small in the surface of that
droplet.

\subsection{Thermodynamic relation.}
As mentioned in the introduction, the original context of the
fluctuation theorem concerned the large deviations in the entropy
production rate of a dynamical system.  Since we have not specified
any dynamics here, we must postpone a related discussion to the
next section.  Yet, we can compare with the thermodynamic
potentials.

To start define the energy function
\begin{equation}
\Phi_0(U) \equiv \sum_{A\ni 0} \frac{U_A}{|A|}
\end{equation}
and its translations $\Phi_x(U)(\eta) = \Phi_0(U)(\tau_x \eta)$. We
define the free energy density for the interaction $U$ as
\begin{equation}
P(U) \equiv \lim_n \frac 1{|\Lambda_n|} \ln
\sum_{\sigma\in \Omega_{\Lambda_n}} \exp[-\beta\sum_{A \subset
\Lambda_n} U_A(\sigma)]).
\end{equation}
This coincides with
\begin{equation}
P(U) = \lim_n \frac 1{|\Lambda_n|} \ln Z_{\Lambda_n}^\beta(\eta)
\end{equation}
of (\ref{pfie}) for all boundary conditions $\eta$.

Finally, the entropy density of a translation-invariant probability
measure $\mu$ is
\begin{equation}\label{enden}
s(\mu) \equiv - \lim_n \frac 1 {|\Lambda_n|}
\sum_{\sigma\in\Omega_{\Lambda_n}} \mu_n[\sigma] \ln \mu_n[\sigma]\geq
0
\end{equation}
where $\mu_n[\sigma]$ is the probability for the measure $\mu$ to
find the configuration $\sigma$ in the box $\Lambda_n$ (and $0\ln 0
= 0$). The relative entropy density between two translation-invariant
probability measures $\mu$ and $\rho$ (with $\rho_n(\sigma)
=0$ implying $\mu_n(\sigma)=0$) is
\begin{equation}\label{endenrel}
s(\mu|\rho) \equiv
\lim_n \frac 1 {|\Lambda_n|}
\sum_{\sigma\in\Omega_{\Lambda_n}} \mu_n[\sigma] \ln
\frac{\mu_n[\sigma]}{\rho_n[\sigma]} \geq 0.
\end{equation}
\\
If $\mu$ is a translation-invariant Gibbs
measure (at inverse temperature $\beta$) for the interaction $U$,
then
\begin{equation}\label{var}
P(U) = s(\mu) - \beta\mu(\Phi_0(U)).
\end{equation}
For a given interaction $(U_A)$ we also like to have around the
free energy functional $F(U,\rho)$ defined for
translation-invariant probability measures $\rho$ by
\begin{equation}
F(U,\rho) \equiv s(\rho) -\beta\rho(\Phi_0(U)).
\end{equation}
We have, besides $F(U,\mu)=P(U)$ for the Gibbs measures $\mu$ with
respect to $U$, see (\ref{var}), that
\begin{equation}
P(U) > F(U,\rho)
\end{equation}
for all translation-invariant probability measures $\rho$ which are
not Gibbs measures for $U$ at inverse temperature $\beta$ (Gibbs'
variational principle).

The $\pi-$ transformed interaction potential $\pi U$ is defined via
\begin{equation}
\pi U_A = \pi_{n(A)}U_A
\end{equation}
and the $\pi-$ transformed measure $\pi \mu$ is obtained by its
expectations for all local functions $f$:
\begin{equation}
\pi \mu(f) = \mu \circ \pi_n (f)
\end{equation}
for $n=n(f)$ so that $D_f\subset \Lambda_n$.
 Clearly, $P(\pi U) =
P(U)$ by the assumed $\pi-$ invariance of the counting measure, see
(\ref{trans}). (This also follows from observing that $-P(\pi U) +
P(U)= p(2E,E)=p(0,E)=0$ by (\ref{flu}) and (\ref{free}).) For the
same reason, $s(\pi \mu) = s(\mu)$ and if $\mu$ is a Gibbs measure
for $U$, then $\pi \mu$ is a Gibbs measure for $\pi U$ (and vice
versa). (To avoid trivialities, it is understood that the
interaction $\pi U$ is not physically equivalent with $U$ as long
as some $E^\alpha\neq 0$.)

 We next show that the averaged current (whose fluctuations are investigated in
Theorem \ref{thm:cg}) is always (strictly) positive as it equals a
relative entropy density.  To link it also to a free energy
production we must require that the free energy $P(U + t(\pi U-U))$
is differentiable with respect to $t$ at $t=0$. For this (see e.g.
\cite{Geo}), it suffices e.g. that
\begin{equation}\label{assu}
\sum_{A\ni 0} |A| ||U_A^0|| , \sum_{\alpha=1}^m E^\alpha
\sum_{A\ni 0} |A| ||U_A^\alpha|| <
1.
\end{equation}

\begin{prop}\label{prop:en}
 For the Gibbs measure $\mu\neq \pi \mu$ of Theorem \ref{thm:cg},
\begin{equation}\label{onsa}
s(\pi\mu|\mu) = s(\mu|\pi \mu) = 2\beta \sum_{\alpha=1}^m E^\alpha
\mu(J_0^\alpha)
> 0
\end{equation}
and, under the assumption (\ref{assu}), is also given via
\begin{equation}\label{ons}
\sum_{\alpha=1}^m E^\alpha \mu(J_0^\alpha) =
- \frac 1{2\beta} \frac{\partial}{\partial t} P(U + t(\pi U-U))(t=0).
\end{equation}
\end{prop}
\medskip
\noindent
{\bf Proof:} The positivity follows from the variational principle:
\begin{equation}\label{pcause}
s(\mu) -\beta \mu(\Phi_0(U)) = P(U) > F(U,\pi \mu) = s(\mu)
-\beta \pi\mu(\Phi_0(U)).
\end{equation}
As is well known the relative entropy (\ref{endenrel}) can be
rewritten as a difference of free energies: $s(\pi\mu|\mu) = P(U)
- F(U,\pi\mu)$.  We can now use that
\begin{equation}
\beta \mu(\Phi_0(\pi U)-\Phi_0(U)) =
- \frac{\partial}{\partial t} P(U + t(\pi U-U))(t=0)
\end{equation}
is exactly equal to $2\beta
\sum_{\alpha=1}^m E^\alpha \mu(J_0^\alpha)$, as required.$\Cox$
\medskip

\noindent
{\bf Remark 1:} The positivity of (\ref{onsa}) should be compared
with (\ref{popo}).  The positivity of the entropy production is
discussed in \cite{Ru1,Ru2}.  The positivity of (\ref{onsa}) just
follows here from the Gibbs' variational principle: with $\dot{s}_n
\equiv \sum E^\alpha \sum_{x\in \Lambda_n} J_x^\alpha$, for $\mu-$
almost every $\sigma, \dot{s}_n(\sigma)/|\Lambda_n| \rightarrow
s(\mu|\pi\mu)/2\beta > 0$ where the almost sure convergence assumes
that $\mu$ is a phase.  That $s(\mu|\pi\mu)$ has something to do
with entropy production will become clear in the next section when
a dynamics and the time-reversal operation is considered.

\noindent
{\bf Remark 2:} Thinking about $s(\mu|\pi\mu)$ as entropy
production, (\ref{onsa}) gives the usual bilinear expression in
terms of thermodynamic fluxes and forces.  Remember that the
dependence of $p(\lambda,E)$ on $E$ in (\ref{free}) comes from the
state $\mu$.  The $E^\alpha$ correspond to field strengths or
amplitudes producing energy- or particle flow.  Of course, on the
formal level above, the distinction must remain arbitrary and one
can of course include the $E^\alpha$ in the potentials
$U_A^\alpha$.

\subsection{Green-Kubo formula.}
It has been observed in other places, \cite{K,LS,G}, that the
fluctuation theorem quite directly gives rise to various familiar
formulae of linear response.  We will not pursue this matter here
very far except for repeating the simplest derivations.

Assuming smoothness of the free energy in the external fields, we
differentiate (\ref{flu}) with respect to $E^\gamma$ and
$\lambda_\alpha,
\alpha,\gamma=1,\ldots,m$ at $E=\lambda=0$:
\begin{equation}
\frac{\partial}{\partial E^\gamma}\frac{\partial}{\partial
\lambda_\alpha} p (0,0) = -
\frac{\partial}{\partial E^\gamma}\frac{\partial}{\partial
\lambda_\alpha} p (0,0) -2
\frac{\partial}{\partial \lambda_\gamma}\frac{\partial}{\partial
\lambda_\alpha} p (0,0).
\end{equation}
On the other hand,
\begin{equation}
\frac{\partial}{\partial \lambda_\alpha} p(0,E) = \beta
\mu(J^\alpha_0)
\end{equation}
while
\begin{equation}
\frac{\partial}{\partial \lambda_\gamma}\frac{\partial}{\partial
\lambda_\alpha} p (0,0)= -\beta^2 \sum_x \nu(J^\alpha_0
J_x^\gamma).
\end{equation}
Conclusion,
\begin{equation}\label{GK}
\frac{\partial}{\partial E^\gamma} \mu(J_0^\alpha)(E=0) =
\beta \sum_x \nu(J^\alpha_0
J_x^\gamma),
\end{equation}
and the change in relative entropy $s(\mu|\pi\mu)$ (see
(\ref{onsa})) from Proposition \ref{prop:en} is in quadratic
approximation for small $E$ given by
\begin{equation}\label{on}
s(\mu|\pi\mu) = 2\beta^2 \sum_{\alpha,\gamma} E^\alpha E^\gamma
\sum_x \nu(J^\alpha_0 J_x^\gamma).
\end{equation}
Equation (\ref{GK}) is a standard Green-Kubo relation while
(\ref{on}) expresses the relative entropy density $s(\mu|\pi\mu)$
(or change in free energy) in terms of the current-current
correlations (with the obvious analogues of Onsager symmetries). In
conclusion, we have identified a (model-dependent) continuous
function
\begin{equation}
\dot{S}(\sigma) \equiv \sum_\alpha E^\alpha J_0^\alpha(\sigma)
\end{equation}
with $\pi \dot{S} = - \dot{S}, \mu(\dot{S}) > 0, \nu(\dot{S}) = 0$
and symmetric response matrix
\begin{equation}
\frac{\partial}{\partial E^\gamma} \mu(\frac{\partial}{\partial
E^\alpha} \dot{S})(E=0) = \beta \sum_x \nu(J^\alpha_0 J_x^\gamma).
\end{equation}
Symmetries in higher order terms can be obtained by taking higher
derivatives of the generating formula (\ref{flu}).\\  The notation
$\dot{S}$ should not be read as a time-derivative (change of
entropy in time).  More appropriate will be to regard
$\mu(\dot{S})/2$ as the limit $(S_f - S_i)/T$ as time $T$ goes to
infinity of the total change of entropy $S_f - S_i$ in a reservoir
during the nonequilibrium process. The reservoir is initially in
equilibrium with thermodynamic entropy $S_i$ and after absorbing
the heat dissipated by the nonequilibrium process it reaches a new
equilibrium with entropy $S_f$.  We will come back to this once
time has been explicitly introduced (in Section 3).

\section{Fluctuation theorem for PCA.}

PCA (short for probabilistic cellular automata) are discrete time
parallel updating stochastic dynamics for lattice spin systems.
They are used in many contexts but we see them here as interesting
examples of non-equilibrium dynamics. We refer to \cite{Gol,LMS}
for details and examples and we restrict ourselves here to the
essentials we need. We work with time-homogeneous
translation-invariant nearest-neighbor PCA which are specified by
giving the single-site transition probabilities
\begin{equation}\label{trapro}
0 < p_{i,n}(a|\sigma) = p_i(a|\sigma(j,n-1), |j-i|\leq 1) < 1, a\in
S, \sigma \in \Omega_{d+1}.
\end{equation}
This defines a Markov process $(X_n)_{n=0,1\ldots}$ on $\Omega_d$
for which for all finite $V\subset \Z^d$,
\begin{equation}
\mbox{ Prob }[X_n(i)=a_i, \forall i \in V|X_{n-1}]=
\prod_{i\in V} p_i(a_i|X_{n-1}(j), |i-j|\leq 1), a_i\in S
\end{equation}
with some given initial configuration $X_0=\xi \in \Omega_d$.
 Notice that we have
kept the notation $\sigma$ for a general configuration on the
space-time lattice.  Remember that $x=(i,n)\in \Z^{d+1}$ stands for
a space-time point with time-coordinate $n$ at site $i\in \Z^d$.
\\The $\pi_\Lambda$ are restricted to
time-reversal transformations and the volumes $\Lambda$ are to grow
first in the time-direction (for a fixed spatial window).

\subsection{Steady state fluctuation theorem.}
If we take a translation-invariant stationary state $\rho$ of a PCA
as above, then its Markov extension defines a translation-invariant
Gibbs measure $\mu$ for the (formal) Hamiltonian
\begin{equation}\label{pcaen}
H(\sigma) = - \sum_{i,n} \ln
p_{i,n}(\sigma(i,n)|\sigma(\cdot,n-1)).
\end{equation}
We refer to \cite{Gol,LMS} for a precise formulation.  $\mu$
describes the distribution of the space-time configurations in the
steady state and its restriction to any spatial layer is equal to
the stationary state $\rho$ we started from.  To characterize
$\rho$, one must study the projection of $\mu$ to a layer (see
\cite{MRV} for a variational characterization of such a
projection).\\
Since $\mu$ is Gibbsian we can try applying the
theory of the previous section. Most interesting is to consider a
sequence of rectangular boxes $\Lambda_{L,N}
\equiv
\{x=(i,n) \in
\Z^{d+1}=\Z^d\times \Z: |i|
\leq L, |n| \leq N\}$.  The idea is that we wish to keep the
spatial size $L$ much smaller than the time-extension $N>>L$. As
transformation we take $\pi_{\Lambda_{L,N}} \equiv \pi_{L,N}$
corresponding to a time-reversal:
\begin{equation}
\pi_{L,N}\sigma(i,n) = \sigma(i,-n), |n|\leq N, |i|\leq L
\end{equation}
and $\pi_{L,N}\sigma(i,n) =
\sigma(i,n)$ whenever $(i,n)\notin \Lambda_{L,N}$.

Define the current
\begin{equation}\label{pcacur}
J_{i,n}(\sigma) \equiv
\ln p_i(\sigma(i,n)|\sigma(\cdot,n-1)) -
\ln p_i(\sigma(i,n-1)|\sigma(\cdot,n)).
\end{equation}
Notice that in contrast with the previous section, we do not
specify here the unperturbed state (but one can always take some
homogeneous product measure) and we take $m=1=2E$ for simplicity.
$J_{i,n}$ is a local function and it is the space-time translate of
$J_0$. In the same way as in (\ref{rele}), we define
\begin{equation}
R_{L,N}(\sigma) \equiv H(\pi_{L,N}\sigma) -H(\sigma).
\end{equation}
Starting from (\ref{pcaen})  $R_{L,N}$ can be written out as a
finite sum but most important is that
\begin{equation}\label{es1}
R_{L,N}(\sigma) = \sum_{n=-N+1}^{N-1} \sum_{|i|\leq L-1}
J_{i,n}(\sigma) + G_{L,N}(\sigma),
\end{equation}
where
 \begin{equation}\label{es2}
\|G_{L,N}\| \leq c(2N+1)(2L+1)^{d-1} + c'(2L+1)^d \leq c(d) N L^{d-1},
\end{equation}
with a constant $c(d)$ depending on the dimension
$d$ and on the transition probabilities (\ref{trapro}).  We are
therefore in a position to repeat the fluctuation Theorem
\ref{thm:cg} in that context.

\begin{thm}\label{thm:cgpca}
Take $L=L(N)\leq N$ growing to infinity as $N\uparrow \infty$. The
limit
\begin{equation}
e(\lambda) \equiv -\lim_N \frac 1{|\Lambda_{L,N}|}
\ln \mu(\exp[-\lambda \sum_{x\in \Lambda_{L-1,N-1}} J_{i,n}])
\end{equation}
exists for all real $\lambda$ and
\begin{equation}\label{eee}
e(\lambda) = e(1-\lambda).
\end{equation}
Moreover, for fixed $L$,
\begin{equation}
e_{L,N}(\lambda) \equiv
- \frac 1{N}
\ln \mu(\exp[-\lambda \sum_{x\in \Lambda_{L-1,N-1}} J_{i,n}])
\end{equation}
(which, generally, is of order $L^d$) satisfies
\begin{equation}
|e_{L,N}(\lambda) - e_{L,N}(1-\lambda)| \leq c(\lambda,d) L^{d-1}
\end{equation}
uniformly in $N\uparrow \infty$.
\end{thm}
\medskip
\noindent
{\bf Proof:} The proof is a copy of the proof of Theorem
\ref{thm:cg}.  As before, we have automatically, from the Gibbs
property (as in (\ref{1.14})), that
\begin{equation}
\mu(\exp[-\lambda R_{L,N}]) = \mu(\exp[-(1-\lambda)R_{L,N}]).
\end{equation}
We now substitute (\ref{es1}) and use the estimate (\ref{es2}) to
perform the limits.$\Cox$
\medskip

\noindent
{\bf Remark 1:} One may wonder about the existence of the limit
$e_L(\lambda)\equiv\lim_N e_{L,N}(\lambda)$ for fixed $L$. This is
certainly expected when the steady-state $\mu$ is a high
temperature Gibbs state.  In that case, the limit $\lim_L
e_L(\lambda)/L^d
= \lim_L e_L(1-\lambda)/L^d$ satisfies (\ref{eee}).

\noindent
{\bf Remark 2:} Some quite similar results were discussed already
in \cite{G3}.  There however the dynamics was deterministic (weakly
coupled strongly chaotic maps).  There again, the methods of
\cite{BK,BK1,PS,PY} can reduce the problem to a higher dimensional
symbolic dynamics and the methods of the previous section are ready
for use.

\subsection{Entropy production.}
The measure $\mu$ gives the probability distribution of the
space-time histories in a steady-state.  It is therefore natural to
consider $s(\mu)$ (see (\ref{enden})) as its specific entropy rate
(i.e, entropy per unit volume and per unit time).  In terms of the
stationary state $\rho$ we have (see \cite{Gol}) that
\begin{equation}\label{rhoent}
s(\mu) = -\rho(\sum_{a\in S} p_0(a|\sigma(\cdot,-1)) \ln
p_0(a|\sigma(\cdot,-1)).
\end{equation}
On the other hand, the free energy density vanishes identically for
PCA (because of the normalization in (\ref{pcaen}), see
\cite{Gol,LMS}), so that, from (\ref{pcause}),
\begin{equation}\label{expli1}
P(U) = 0, P(U)-F(U,\pi \mu) = -s(\mu)
 + \beta \pi\mu(\Phi_0(U)).
\end{equation}
Hence, still in the notation of the previous section, whenever
$P(U)=0$ (which is verified for PCA),
\begin{equation}
-s(\mu|\pi\mu) = s(\mu) - \beta \pi\mu(\Phi_0(U)).
\end{equation}
(This formula is not correct when we replace in it $\mu$ by
$\pi\mu$.) That is interesting because we found that now
 $s(\mu|\pi\mu) > 0$
is minus the specific entropy rate $s(\mu)$ modulo a term which is
linear in $\mu$.  Writing this out in our present notation, this is
nothing else than
\begin{equation}\label{expli}
-\mu(J_0) =
-\rho(\sum_{a\in S} p_0(a|\sigma(\cdot,-1)) \ln
p_0(a|\sigma(\cdot,-1)) - \mu(-\ln p_0(\sigma(0)|\sigma(\cdot,1)).
\end{equation}
The first term to the right is the specific entropy rate
(\ref{rhoent}) (always positive) and the second term (linear in
$\mu$) subtracts from this exactly so much that the net-result to
the left vanishes in the case of time-reversal symmetry (detailed
balance). Of course, as in Proposition \ref{prop:en} we have an
equality between the averaged current in the steady state $\mu$ and
the relative entropy $s(\mu|\pi
\mu)$ (remember that we took $m=1=2E$!).  We can therefore conclude
that indeed $\mu(J_0)$ or $s(\mu|\pi\mu)$ must be regarded as the
(positive) entropy production by our dynamics.  The current
associated to the breaking of time-reversal symmetry gives rise to
nothing else than the local (in space-time) entropy production
whose fluctuations we have investigated in Theorems \ref{thm:cg}
and \ref{thm:cgpca}.  The points made in Section 2.4 related to the
Green-Kubo formula remain unaltered and we do not repeat them
here.\\ Yet, to obtain a physically inspiring picture, we should
connect the above analysis to measurable quantities.  The (second
part of the) second law of thermodynamics connects the
thermodynamic entropy of an initial and final equilibrium state
after some thermodynamically irreversible process has taken place.
In an adiabatic non-quasi-stationary process the entropy can only
increase: $S_f > S_i$.  If we now were to rerun the process in the
opposite direction, simply by (thermodynamically) inverting all the
currents (by changing the sign of all  gradients of the intensive
variables), again the entropy would increase and by the same amount
as before ($s(\mu) = s(\pi\mu)$) and we would reach a new
equilibrium with entropy equal to $S_i + 2(S_f - S_i)$. While we
lack at this point a more formal understanding, we believe that our
entropy production exactly measures that difference: $[S_f - S_i]
- [S_i - S_f] = 2(S_f - S_i)
= s(\mu|\pi\mu)
> 0$.  More generally and depending on the physical realization of the process, these
considerations must apply to the relevant thermodynamic potential
and `entropy production' must for example be replaced by `work
done' or `free energy production.'\\
 We will further illustrate this by an example in the following
subsection but it is interesting to remark already that
$s(\mu|\pi\mu)$ reproduces, via the formal analogies on the level
of the variational principle (both for Gibbs and for SRB states),
the entropy production in the context of the theory of dynamical
systems. There we have that the entropy production is given by
(\ref{popo}) with $\rho(\dot{S})$ equal to the sum of the positive
Lyapunov exponents with respect to $\phi^{-1}$ minus the sum of
positive Lyapunov exponents with respect to $\phi$.  If $\rho$ is
singular with respect to $d\xi$ and has no vanishing Lyapunov
exponent, then $\rho(\dot{S}) > 0$, see \cite{Ru1}. In our case,
$s(\mu|\pi\mu) =
\beta\pi\mu(\Phi_0(U)) - \beta\mu(\Phi_0(U)) = P(U) - F(U,\pi\mu)
> 0$.

\subsection{Illustration.}

We take here a closer look at the current (\ref{pcacur}) for Markov
chains.  The spatial degree of freedom $i\in \Z^d$ has now
disappeared and we must study
\begin{equation}\label{macur}
J_n(\sigma) \equiv \ln p(\sigma(n)|\sigma(n-1)) - \ln
p(\sigma(n-1)|\sigma(n))
\end{equation}
for $\sigma \in \Omega_1$ and transition probabilities
\begin{equation}
\mbox{ Prob }[X_n=a|X_{n-1}=b] = p(a|b), a,b\in S
\end{equation}
for the stationary $S$-valued Markov chain $X_n$. The steady state
$\mu$ is now a homogeneous one-dimensional Gibbs measure and its
single-time restriction is the stationary measure $\rho$  on $S:
\sum_b p(a|b)\rho(b) = \rho(a), a\in S$.

The steady state expectation of the current (\ref{macur}) is
\begin{equation}\label{ppp}
\mu(J) = \sum_b\rho(b) \sum_a p(a|b)[\ln p(a|b)-\ln p(b|a)].
\end{equation}
Now use that the transition probabilities $q(\cdot|\cdot)$ for the
reversed chain $(Y_n\equiv X_{-n})_n$ (with distribution $\pi
\mu$ but with the same stationary measure $\rho$) are given by
\begin{equation}\label{revers}
q(a|b) \equiv \mbox{ Prob }[X_n=a|X_{n+1}=b] =
p(b|a)\frac{\rho(a)}{\rho(b)}.
\end{equation}
Since $\sum_b \rho(b) \sum_a p(a|b) \ln \rho(b) = \sum_b \rho(b)
\ln \rho(b) = \sum_b \rho(b) \sum_a p(a|b) \ln \rho(a)$, we can
substitute (\ref{revers}) into (\ref{ppp}) ($q(a|b)$ for $p(b|a)$)
with no extra cost and we obtain
\begin{equation}\label{best}
\mu(J_0) = \rho(S(p|q))
\end{equation}
where
\begin{equation}\label{marel}
S(p|q) = \sum_a p(a|\cdot) \ln \frac{p(a|\cdot)}{q(a|\cdot)} \geq
0.
\end{equation}
is the relative entropy between the forward and the backward
transition probabilities.  (\ref{marel}) is zero only if the Markov
chain is time-reversible (in which case $\mu=\pi\mu$). Then,
(\ref{revers}) for $q(a|b)=p(a|b)$ becomes the detailed balance
condition. Relation (\ref{best}) is nothing but (\ref{onsa})
specified to the context of Markov chains.

A second less trivial and physically interesting illustration can
be taken from a model of hopping conductivity.  It is a bulk driven
diffusive lattice gas where charged particles, subject to an
on-site exclusion, hop on a ring in the presence of an electric
field. The configuration space is $\Omega_1= \{0,1\}^{\cal T}$ with
$\xi(i)=0$ or $1$ depending on whether the site $i\in {\cal T}$ is
empty or occupied. We take for ${\cal T}$ the set
$\{1,\ldots,\ell\}$ with periodic boundary conditions.  To each
bond $(i,i+1)$ in the ring and independently of all the rest there
is associated a Poisson clock (with rate 1). If the clock rings and
$\xi(i)=1,\xi(i+1)=0$ then the particle at $i$ jumps to $i+1$ with
probability $p$.  If on the other hand, $\xi(i)=0,\xi(i+1)=1$ the
particle jumps to $i$ with probability $q$.  Therefore, the
`probability per unit time' to make the transition from $\xi$ to
$\xi^{i,i+1}$ (in which the occupations of $i$ and $i+1$ are
interchanged) is given by the exchange rate
\begin{equation}
c(i,i+1,\xi) = p\xi(i)(1-\xi(i+1)) + q\xi(i+1)(1-\xi(i)).
\end{equation}
and should be thought of as a continuous time analogue of
(\ref{trapro}).  It is natural to call $E
=
\ln p/q$ the electric field. This model is called the asymmetric
simple exclusion process and it is also considered in \cite{LS}.
Strictly speaking, it is not a PCA but a continuous time process
with sequential updating.  However, since it is a jump process, the
change with respect to the PCA of above just amounts to randomizing
the time between successive transitions.\\
 Each uniform product
measure $\rho$ is time-invariant for this process and we consider
the steady state $\mu$ starting in this invariant state. If we now
consider a realization $\sigma$ of the process in which at a
certain time, when the configuration is $\xi \in \Omega_1$, a
particle hops from site $i$ to $i+1$, then the time-reversed
trajectory shows a particle jumping from $i+1$ to $i$.  The
contribution of this event to the entropy production is therefore
\begin{equation}\label{asym}
\ln c(i,i+1,\xi) -\ln c(i,i+1,\xi^{i,i+1}) =
E[\xi(i)(1-\xi(i+1))- \xi(i+1)(1-\xi(i))].
\end{equation}
This formula is the continuous time analogue of (\ref{macur}) or
(\ref{pcacur}) (but we do not take $E=1/2$ here) with $\xi$ the
configuration right before the jump and $\xi^{i,i+1}$ the
configuration right after the jump in the trajectory $\sigma$. Of
course, this jump in $\sigma$ itself happens with a rate
$c(i,i+1,\xi)$.  We see therefore that the derivative of
(\ref{asym}) with respect to $E$ has expectation
\begin{equation}\label{asymcur}
\mu(J_{i,t}) = \rho(c(i,i+1,\xi) [\xi(i)(1-\xi(i+1))-
\xi(i+1)(1-\xi(i))]) = (p-q) u(1-u)
\end{equation}
for $u\equiv \rho(\xi(i))$ the density.  (\ref{asymcur}) is indeed
the current as it appears in the hydrodynamic equation, here the
Burgers equation, through which a density profile evolves. The
fluctuations of the particle current satisfy (\ref{flu}) or
(\ref{eee}) (with $E=1/2$), see also \cite{LS}. The entropy
production (as in (\ref{best})-(\ref{marel})) is
\begin{equation}
\frac 1{2} s(\mu|\pi\mu) =
\rho(c(i,i+1,\xi)\ln\frac{c(i,i+1,\xi)}{c(i,i+1,\xi^{i,i+1})})=
E (p-q) u(1-u)
\end{equation}
which is the field times the current and is left invariant by
changing $E$ into $-E$. If, to be specific, we take
$p=1/(1+e^{-E})=1-q$, then, in quadratic approximation,
\begin{equation}
\frac 1{2} s(\mu|\pi\mu) = u(1-u) E^2
\end{equation}
which is the dissipated heat through a conductor in an electric
field $E$ with Ohmic conductivity $u(1-u) = \mu(J_0^2)(E=0) =
\rho(c(0,1,\xi)[\xi(0)(1-\xi(1)) - \xi(1)(1-\xi(0))]^2)$
given by the variance of the current.  This model (together with
the models discussed in \cite{LS}) illustrates that the methods
exposed in the present paper are not restricted to just PCA.  We
have restricted us here to a somewhat informal treatment of the
aspects concerning the entropy production in the model as it will
be included in a future publication dealing with the local
fluctuation theorem, \cite{M2}.

\section{Concluding remark.}

It does not seem unreasonable that Gibbs' variational principle
determining the conditions of equilibrium can be generalized to
certain nonequilibrium conditions.  In this note we have shown that
describing the steady state via the standard methods of the Gibbs
formalism leads directly to the fluctuation theorem.  This is true
close or far from equilibrium because it follows quite generally
from the defining Gibbs property itself. From this `Gibbsian' point
of view, applying the local fluctuation theorem to various specific
models is to add specific observable consequences to the studies of
E.T. Jaynes, \cite{Ja}.

\end{document}